# Insights on drying and precipitation dynamics of respiratory droplets in the perspective of Covid-19

[1]Saptarshi Basu , [1]Prasenjit Kabi,  [2]Swetaprovo Chaudhuri and [3]Abhishek Saha

[2]*Department of Mechanical Engineering, Indian Institute of Science, Bengaluru, India2*[3]*Institute for Aerospace Studies, University of Toronto, Toronto, Canada*

[3]*Department of Mechanical and Aerospace Engineering, University of California San Diego, La Jolla, USA*

*We isolate a nano-colloidal droplet of surrogate mucosalivary fluid to gain fundamental insights into the infectivity of air borne nuclei and viral load distribution during the Covid-19 pandemic. Saltwater solution containing particles at reported viral loads are acoustically trapped in contactless environment to emulate the drying, flow and precipitation dynamics of real airborne droplets. Observations with the surrogate fluid are validated by similar experiments with actual samples from a healthy human subject. A unique feature emerges with regards to the final crystallite dimension; it is always 20-30% of the initial droplet diameter for different sizes and ambient conditions. Airborne-precipitates nearly enclose the complete viral load within its bulk while the substrate precipitates exhibit a high percentage (~80-90%) of exposed virions (depending on the surface). The letter demonstrates the leveraging of an inert nano-colloidal system to gain insights into an equivalent biological system.*

Humans routinely eject pulsatile jets containing microdroplets [1,2] during sneezing, coughing or even talking, which aid in rapid transport of viral loads[3] leading to pandemics such as the Covid-19[4,5]. Such droplets remain airborne for considerable amount of time, given the initial size and ambient conditions, and evaporate to form infective nuclei[6].. Chaudhuri et al[7] elucidated the mechanics by which droplet initiates and propagates a pandemic by combining models of droplet evaporation, aerodynamics and SIR[7]. Given the size distribution of respiratory droplets[8], the airborne nuclei have a high probability of assimilation via oral or nasal passage. They might also deposit on objects of daily use to form fomites which can subsequently be assimilated by a person via touch. Although the infectivity of a given droplet-nuclei/fomite is linked to the initial viral load[9,10] as well its stability in different environments[11,12,13], it is equally important to understand the desiccation and the precipitation dynamics of the infected droplet. The general practice is to study the viral activity in cellular environments[14] under diffusion effects [15,16] where the precipitation dynamics are not very important. On the other hand, droplet embodies a plethora of fluidic transport[17,18] and couples precipitation and evaporation of droplet to the agglomeration dynamics of the virions with the cellular material to which it is attached. Given the complexity of the experiment with actual viruses in respiratory fluid, such studies have been rarely attempted[20]. Mucosalivary fluids are known to have dissolved salts (~1 wt.%) in addition to mucus and enzymes[19,20]. This letter uses dissolved NaCl in de-ionised water at 1 wt% as a simple surrogate liquid. Inactive nanoparticles of polystyrene (mean size 100 nm) are added to this saline solution to emulate virus particle of similar size (CoV-2, Influenza)[21] Although having different mechanical and chemical properties, nanoparticles shed important insights into the transport and precipitation role of virions inside an airborne droplet. Viral loads occur in the range of $10^6$-$10^9$ per ml of the respiratory fluid[22] which translates into an approximate initial nanoparticles load of $\varphi_{np} \sim 0.0001$ (in wt. % unless stated otherwise) in the given saline solution. However, precipitation dynamics at higher loads[23] presents a fundamental insight into nanoparticles interaction at high electrolyte concentration[24] as well as a crucial premise for several other applications[25]. To this end, $\varphi_{np}$ would also be varied from 0.01 to 0.1 for further investigation.

Given the experimental complexity associated with studying a mobile air-borne droplet, we have used an acoustic levitator to trap a droplet in the air (tec5) and allowed it to evaporate in a controlled ambience ($T_\infty$=28$\pm$0.2 °C, $RH_\infty$=41$\pm$2%). Acoustic levitation[26] has been extensively used to study the evaporation[27] and precipitation dynamics of a solute laden droplet[25,28,29,30]. A droplet of the surrogate fluid having an initial diameter $D_0$=550 μm+10 μm is inserted into one of the stable nodes of the acoustic levitator and imaged every 3 seconds at 30 frames per second (see S1 of Supporting Information) till the end of evaporation. The effective diameter of the droplet $D = \sqrt[3]{d_x^2 d_y}$ where $d_x$ and $d_y$ are the major and minor axis of the droplet, respectively. Figure 1b shows the lifetime of evaporation. The droplet monotonically reduces till the time instant $t=t_I$ where the shrinkage appears arrested. Subsequently, the shape of the droplet deviates from its initial sphericity ($d_x/d_y$=1) at $t=t_{II}$ and finally assumes its crystalline form at $t=t_{III}$ shown for different concentrations of nanoparticles.

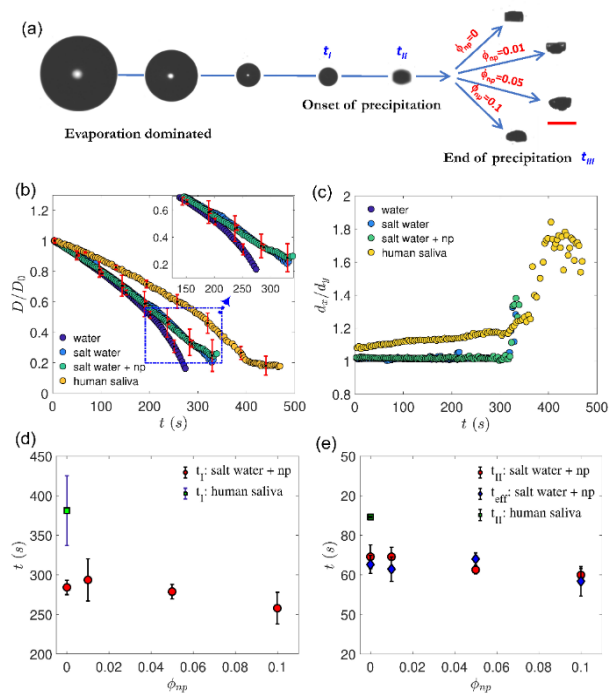

**Figure 1 Evaporation dynamics of a levitated droplet. (a) Sequential snapshots show the reduction of droplet diameter culminating into the final precipitate shown for different values of nanoparticle concentration ($\varphi_{np}$ in wt.%). Time instant $t_I$ indicates the end of evaporation dominated stage when the rate**

of diameter reduction reduces sufficiently. Time instant $t_{II}$ indicates the departure from sphericity of the droplet. The time instant $t_{III}$ indicates the end of the process. The scale bar is 0.2 mm. (b) The droplet diameter is plotted as $D/D_0$ vs. time (t) pure water, salt-water (1 wt. %) and salt-water+nanoparticles(np) where the mean concentration of the range $\varphi_{np}$=0.01-0.1 is used. The mean value of $D_0$=550$\pm$10 μm. Error bars are standard deviations of multiple runs. (c) Aspect ratio of the droplet ($d_x/d_y$) vs. $t$ for the same conditions as (b) where $d_x$ and $d_y$ refer to the major and the minor axis of the droplet respectively. (d) Variation in $t_I$ vs. $\varphi_{np}$ for both surrogate and HS. (e) Comparison of $t_{II}$ for both surrogate and HS. The onset of efflorescence ($t_{eff}$) for different $\varphi_{np}$ is also plotted. Ambient temperature is set at 28$\pm$0.2 °C and the RH is set at 41$\pm$2%.

The diameter reduction of the surrogate fluid (salt-water +nanoparticle) is plotted in Figure 1b. Since the presence of nanoparticles till $\varphi_{np}$=0.1 shows no distinctive effect on the reduction of the dimeter, only the mean concentration ($\varphi_{np}$=0.05) is plotted (Figure 1b). The initial stage of evaporation is diffusion limited[27], fits the standard $D^2$ law which states that[31] $D(t)^2 = D_0^2 - K_e t$. The value of $K_e \sim O$ ($10^{-9}$) m²/s for a pure water droplets and predicts the total lifetime to be $t_{evap} = D_0^2 / K_e \approx 300$ s, which is close to observed values (Figure 1b). Initial droplet reduction rates are nearly equal for water and surrogate fluid but start deviating at $t$>200 s (inset of Figure 1b) due to the presence of dissolved salt which reduces the vapour pressure droplet[25]. This is consistent with the evaporation-precipitation model presented in Chaudhuri et al[7]. The deviation between the surrogate and HS droplets originates in the complex composition of the later (mucus, surfactants, polyelectrolytes, etc) as well as inherent inhomogenity in the sample due to collection methodology. Nonetheless, the $D/D_0$ appears to follow a similar trend at an offset rate and exhibits similar phenomenology.

The end of evaporation dominated phase occurs at $t=t_I$ when the diameter shrinkage dramatically reduces leading to a knee like appearance (see Figure 1b). However, solvent loss, though slower, continues till $t_{III}$. The knee-transition occurs at $t=t_I$=260~300 s for the surrogate droplet and $t_I$=380 s for the HS droplet as shown in Figure 1d. The knee formation is universally observed for both HS and at 0.2~0.3$D_0$ as corroborated from experiments with different initial droplet sizes (300 to 800 μm), temperature range (27-30 °C) and RH (40%-50%) (see S2; Supporting Information). The onset of knee is independent of $\varphi_{np}$ which de-couples viral loading effect on the precipitation dynamics within the respiratory droplet. The distribution of nanoparticles within the droplet bulk can be predicted from the mass Peclet number $Pe_m = \frac{Ur_0}{D_{np}} \sim O(10^2)$, where the appropriate velocity scale, $U$, is the rate of diameter reduction (~2.8 μm/s), $r_0$ is the initial radius of the droplet and $D_{np}$ is the mass diffusivity of nanoparticles in water calculated from the Stokes-Einstein equation $D_{np} = \frac{k_B T}{6\pi\mu r_p} \sim O(10^{-12})$ m²/s. For $Pe_m \gg 1$, the nanoparticles do not diffuse but accumulate near the receding interface of the droplet[24,29].

The droplet shape evolves under evaporation and departs from its initial sphericity as shown by the plot of ($d_x/d_y$) at $t=t_{II}$ (Figure 1c). The transition can be predicted as follows. The Peclet number for a levitated saline droplet[32] is $Pe = \frac{Ur_0}{D_s} \approx 0.5$ where $D_s \sim O(10^{-9})$ m$^2$/s is the diffusion coefficient of NaCl in water[33]. $Pe<1$ indicates homogenous distribution of salt allowing the use of droplet volume to estimate its bulk concentration in the droplet. At a time $t=t_{eff}$ corresponding to $D/D_0<0.26$, the efflorescence limit (640 g/l)[34] is achieved within the bulk of the droplet. The close match between $t_{II}$ and $t_{eff}$ is shown in Figure 1e proving the near coincidence of efflorescence and shape flattening. The HS droplet flattens at an early stage possibly due to the naturally occurring surfactants and the acoustic pressure[19,35] and transitions at $t_{II,HS} \sim 370$ s. Thus, the evaporation and the precipitation dynamics of the surrogate droplet closely match the HS droplet based on the timescales $t_I$ and $t_{II}$. The proposed timescales are well predicted from the evaporation model[7] and simple bulk concentration calculations, indicate the advent of crystallization in saline or HS droplets and applicable toa wide range of droplet size and ambient conditions. Given the absence of particle loading effect on evaporation and precipitation dynamics, only the case of $\varphi_{np}=0$ is used to discuss the role of acoustic streaming on crystallization.

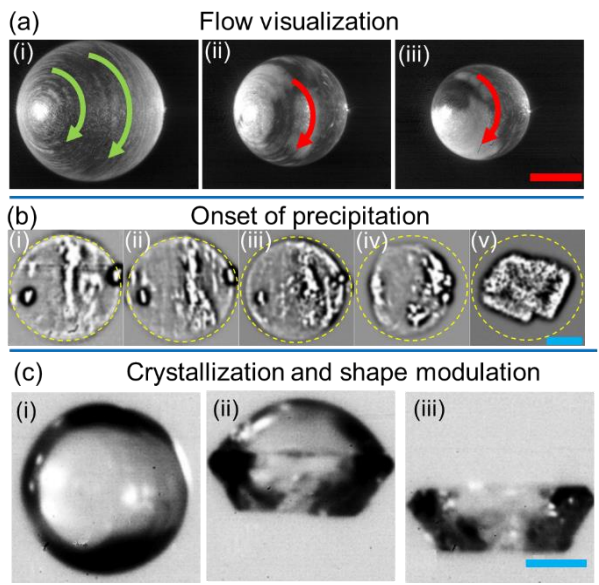

**Figure 2 Flow visualization in case of $\varphi_{np}=0$ is displayed as superposition of three consecutive images (3/2000 s) for $\varphi_{np}=0$ for (i) $D/D_0=1$ (ii) $D/D_0=0.8$ and (iii) $D/D_0=0.7$. The scale bar in red is 0.2 mm. (b) The progression of precipitation in $\varphi_{np}=0$ is laser visualized and presented at (i) $D/D_0=0.27$ (ii) $D/D_0=0.26$, interval (iii) $D/D_0=0.24$ (iv) $D/D_0=0.2$ (v) final crystalline form. (c) Front illuminated droplet shape for $\varphi_{np}=0$ is shown at (i) $D/D_0=0.25$ (ii) Spherical top-half and crystalline bottom half (iii) final crystalline form. The scale bar in blue is 50 μm.**

Acoustic streaming around the droplet governs the internal flow field[26] and is visualized by adding 0.86 μm particles of latex (1.05 g/cc) at an initial concentration of 0.008 % wt. Illumination is done using a laser beam of 1 mm at 0.2 W (see S1 Supporting Information). The time averaged flow field in Figure 2a shows a circulatory motion within the droplet, where a fluid particle near its surface moves at a mean rate of 0.087±0.02 m/s. This homogenizes the salt molecules in the azimuthal direction (but not in the radial direction where it diffuses). The flow magnitude and direction agrees with previous studies of particle image velocimetry in evaporating levitated droplets and remains nearly constant throughout the droplet lifetime[36] as observed from Figures 2a(ii and iii). Note that an ejected respiratory droplet is accompanied by a jet and subjected to atmospheric turbulence leading to similar rotary motions[3] which is recreated in this case due to the acoustic streaming and torque provided by the levitator[26].

Laser scatter in absence of 860 nm particles aid in visualizing the onset of precipitation. The scatter from the droplet is sampled at a rate of 50 fps (for details see S1; Supplementary Information). Images are bandpass filtered to enhance the precipitation induced scatter within the droplet. At $D/D_0$=0.26~0.27, scatter from the centre of the droplet may indicate the onset of precipitation (Figure 2b (i and ii)) which coincides with efflorescence as previously discussed. At $D/D_0$=0.24, the droplet interior shows uniform scatter (Figure 2b(iii)) while the departure from sphericity occurs at $D/D_0$=0.2 (Figure 2b(iv)) which shows an even higher uniformity in scatter. Although, spatial inception of efflorescence is difficult to identify, a drastic shape change could be observed when the bulk has crystallized as seen from the time lapse between Figure 2b (i-iv). The final cuboidal shape of NaCl[37] is observed from Figure 2b(v) at a time $t_{III}$=320~330 s. The shape evolution is better visualized using front illumination (see S1 of Supporting Information) as shown in Figure 2c. The spherical shape in Figure 2c(i) transforms into a dual structure where the lower half has crystallized before the upper half (Figure 2c(ii)). Saha et al[23] attribute this to an unequal pressure distribution at the north and the south poles. Consequently, the salt distribution accumulates faster in the lower half of the droplet leading to earlier crystallization. The final cuboidal shape in Figure 2c(iii) is consistent with Figure 2b(v) but maybe different from those observed from salt precipitation in the atmosphere due to absence of acoustic pressure field. The rate of crystal growth can be estimated as $\frac{0.3D_0 - 0.2D_0}{t_{III} - t_I}$=2~2.3 µm/s. The final crystal dimensions are similar for various nanoparticle loadings (Figure 3a).

The timescales in evaporation and precipitation dynamics is established in the preceding discussions. The morphological similarity between the various precipitates of different compositions ($\varphi_{np}$) (Figure 3a) further demonstrates the independence of precipitation from particle loading rates. To scrutinize the distribution of nanoparticles (emulated viral loading) upon precipitation, marker nanoparticles with fluorescent label (R100, Thermofisher) are loaded into the levitated droplet at $\varphi_{np}$ =0.0001. Precipitation will entrap the nanoparticles in the levitated precipitate (Figure 3b(i)) similar to the entrapment of virions in desiccated airborne droplets. Here z is along the levitator's axis. The preserved levitated precipitate is observed under in fluorescence mode (BX51, Olympus) with a 100x objective (depth of focus ~2.5 µm) at different depths (interval of 3-5 µm). The surface layer in Figure 3b(ii) shows discrete bright spots corresponding to groups of nanoparticles. The image also conveys sparse distribution of nanoparticles. A typical section of the precipitate bulk (Figure 3b(iii) displays diffuse emission from multiple layers, possibly due to higher concentration of particles. Using scanning electron microscopy (SEM), the surface of the same precipitate shown in Figure 3b(ii and iii) is presented in Figure 3b(v). Particles trapped near the upper layers of the precipitate are scanty and appear partially exposed (not stacked as multilayers) as shown in Figure 3b(v). In absence of particle loading, the surface topography of the precipitate is very smooth as shown in Figure 3b(iv). Thus, loaded particles/virions in levitated samples tend to be embedded mostly within the bulk.

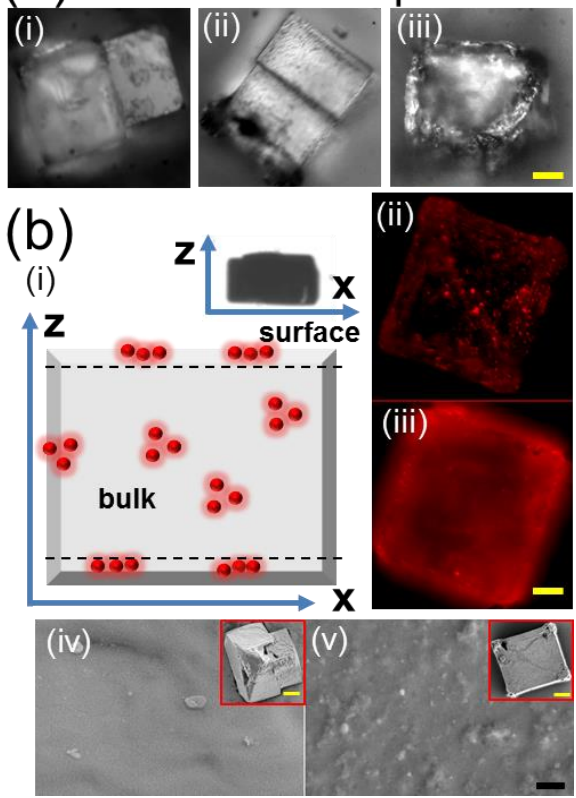

**Figure 3 Micrograph of the preserved precipitate from levitated droplets of (i)$\varphi_{np}$=0 (ii) $\varphi_{np}$=0.1 (iii) $\varphi_{np}$=0.0001(viral load) (b) (i) Schematic of the levitated precipitate with viral load showing entrapped nanoparticles (red spheres). Symbol z represents the levitator axis while x represents the corresponding perpendicular direction. (inset) final shape of the levitated precipitate. (ii) Fluorescent image of the precipitate at depths of (ii) z=1 μm (on surface) (iii) 21 μm (within bulk). (iv) SEM of $\varphi_{np}$=0. (Inset) complete precipitate under SEM. (v) Same sequence of images as (iv) for $\varphi_{np}$=0.0001. Scale bars in yellow equal 20 μm and in black equal 1 μm.**

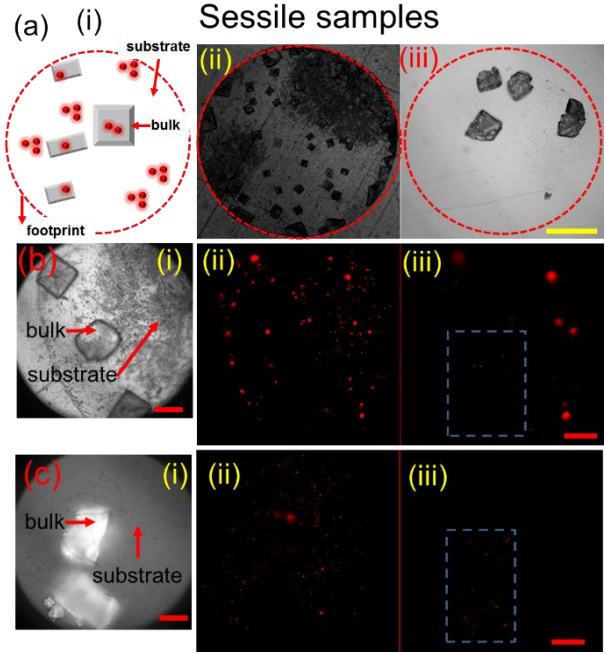

**Figure 4(a) (i) Schematic depicting distribution of nanoparticles within a sessile precipitate. Dashed circle depicts the initial wetted area of the droplet. Complete view of the sessile precipitate on (ii) steel (iii) glass surfaces. (b) (i) Magnified image of the precipitate on steel surface. Fluoresecent images of the same region showing particles exposed on (ii) substrate (iii) and on the bulk of the crystal. (c) Same sequence of images as (b) on glass surface. Scale bar in yellow equals 0.25 mm and in red equals 40 μm.**

While levitated droplets emulate small sized droplets which completely desiccate in air, larger droplets settle to form fomites. Droplets, of same volume and particle loading as the levitated droplets, were dried on steel and glass surfaces. The value of *Pe* >>1 for the sessile case, imply that salt particles will accumulate near the droplet's contact line (Fig. 4a(i)-b(i)). Here, the velocity scale was evaluated based on $U \approx \frac{r}{t_f}$ where $t_f \approx \frac{m_0}{dm/dt} \approx \frac{m_0}{\pi r D_{aw}(1-RH)\rho_v f(\theta)}$, $f(\theta) = 0.27\theta^2 + 1.3$ and the initial value of contact angle $\theta$ (~20° on glass and 60° on steel)[18]. Fluorescence images are acquired within the perimeter of the surface precipitate where the signal from the substrate corresponds to the exposed nanoparticles (Fig. 4b(ii-iii), 4c(ii-iii)). The integrated intensity from the substrate is denoted as $I_{substrate}$ (Fig. 4b(ii), 4c(ii)) while the same from the crystal is $I_{bulk}$ (Fig. 4b(iii), 4c(iii)). The fraction of particles exposed on the surface is simply $I = \frac{\sum I_{substrate}}{I_{total}}$ where $I_{total} = I_{substrate} + I_{bulk}$. Since the given volume of droplet contains n~8 x10$^6$ particles, the average number of particles at the surface is $n_{exp}=nI$. On glass, $n_{exp}$~ 5x10$^6$ while in case of steel substrates, $n_{exp}$~ 7x10$^6$. The variation in the numbers between steel and glass is due to the affinity of the substrate-particle interaction[39] as well as the internal flow structure[40]. Nonetheless, precipitates in fomites indeed show a greater percentage of exposed particles (~ 80-90%) on the substrates as compared to embedded in crystals as in air-borne counterparts. The Probabilistic Analysis for National Threats Hazards and Risks (PANTHR) database[41] predicts the virus lifetime to be significantly shorter (~100 times) in air-borne precipitates when compared to those on solid surfaces.

This correlates to the presented experimental findings that the virions are more exposed in dried settled droplets as opposed to their airborne counterparts.

In summary, a nanocolloidal system is successfully used to mimic the evaporation and precipitation dynamics of an isolated mucosalivary droplet. Theoretical and experimental arguments are presented to show how the evaporation leads to salt crystallization which traps the virion-substitutes at different layers of the air-borne precipitate. Fluorescent microscopy correlates the lower survival rates of virus in the air-borne precipitates to its lower number of exposed virions.

## Acknowledgements

We thank Dr. Sreeparna Majee and Mr. Shubham Sharma for their help during editing the manuscript.